\documentclass[preprint]{jpsj3}
\usepackage{txfonts}
\usepackage[T1]{fontenc} 
\usepackage{graphicx}
\usepackage{dcolumn}
\usepackage{bm}
\usepackage{color}
\usepackage{multirow}
\usepackage{braket}
\usepackage{here} 
\usepackage{indentfirst}
\usepackage{fancyhdr}
\usepackage{ulem}
\title{NMR/NQR and AC-susceptibility Studies in the Weyl Semimetal Superconductor 1T-MoTe$_2$ under Pressure}

\author{T. Fujii$^{1,2}$, H. Yasuoka$^{1}$, M. O. Ajeesh$^{\star}$, M. Schmidt$^{1}$, T. Mito$^{2}$. Yu Liu$^{3,\dagger}$, C. Petrovic$^{3,\ast}$ and M. Baenitz$^{1}$
}

\inst{$^1$Max Planck Institute for Chemical Physics of Solids, 01187 Dresden, Germany¥¥
$^2$Graduate School of Science, University of Hyogo, Kamigori, Hyogo 678-1297, Japan ¥¥
$^3$Brookhaven National Laboratory, Upton, NY 11937-5000, U. S. A.}

\abst{
We performed the Te-nuclear magnetic resonance, the Mo-nuclear quadrupole resonance, and the AC susceptibility in the Weyl semimetal superconductor 1T-MoTe$_2$ at pressures up to 2.17~GPa. From the temperature and pressure dependence of the AC susceptibility, the superconducting transition temperature $T_{\mathrm{c}}$ and the upper critical field $H_{\mathrm{c2}}$ were estimated. The results deviate from the Werthamer-Helfand-Hohenberg model but are well described by $H_{\mathrm{c2}}(T)=H_{\mathrm{c2}}(0)[1-T/T_{\mathrm{c}}]^{\alpha}$. The latter fit yields $H_{\mathrm{c2}}(0)=1.50$~T, $T_{\mathrm{c}}=3.81$K, and $\alpha=1.1$ at 2.17GPa, suggesting that the superconductivity lies in a strong-coupling regime. Since the nuclear spin-lattice relaxation rate divided by temperature, $1/T_1T$, follows the Korringa relation at ambient pressure, the increase in $1/T_1T$ with pressure up to approximately 0.7~GPa indicates an increase in the density of states (DOS), $N(E_\mathrm F)$. This trend mirrors the pressure dependence of $T_{\mathrm{c}}$ in the low-pressure region, consistent with the BCS mechanism. Above 0.7~GPa, however, $N(E_\mathrm F)$ slightly decreases while $T_{\mathrm{c}}$ continues to rise, suggesting an additional pairing contribution beyond the conventional BCS picture. In the 1T$^{\prime}$ phase at 2.17~GPa, the absence of a coherence peak in $1/T_1T$ around $T_{\mathrm c}$, accompanied by a two-step decrease just below $T_{\mathrm c}$, was observed, which may be a signature of unconventional superconductivity.
}
\begin{document}
\maketitle
\thispagestyle{empty}
\section{Introduction}\label{sec1}
In condensed matter physics, the concept of topology has become a fundamental framework for understanding a wide range of quantum materials, including topological insulators and semimetals. Weyl semimetals, a class of topological materials, host the condensed-matter analog of relativistic Weyl fermions that arise from the linear dispersion relation near discrete Weyl nodes in momentum space. These Weyl fermions give rise to exotic physical properties such as surface Fermi arcs~\cite{xu2015, lv2015} and chiral anomaly~\cite{huang2015,zhang2016}. Recently, the Weyl semimetals are also of growing interest for the potential superconductivity hosting Majorana fermions. In this context, 1T-MoTe$_2$ is a promising candidate for a platform to investigate the relationship between the Weyl fermions and the superconductivity.
1T-MoTe$_2$ undergoes a structural phase transition from the high-temperature 1T$^{\prime}$ phase (monoclinic, space group $P2_1/m$) to the low-temperature T$_{\rm d}$ phase (orthorhombic, space group $Pmn2_1$) which is accompanied by the breaking of inversion symmetry and the emergence of a type-II Weyl semimetal state~\cite{sankar2017}. The structural phase transition  occurs at a transition temperature $T_\mathrm t = 250$~K under ambient pressure~\cite{2016YQ,Hu2019a,Takahashi2017,Piva2023,Heikes2018,hang2024}. $T_{\mathrm t}$ decreases linearly with increasing pressure down to 50 $\sim$ 100~K at 0.8~GPa~\cite{Piva2023, Takahashi2017}. Beyond the pressure region near 0.8~GPa where the two phases coexist, the 1T$^{\prime}$ phase becomes dominant at all temperatures above $P$ $\sim 1$~GPa~\cite{Piva2023,Heikes2018}. 

Remarkably, superconductivity emerges in the Weyl semimetal phase of MoTe$_2$ with a transition temperature ($T_{\mathrm c}$) of approximately 0.1~K at ambient pressure~\cite{2016YQ}. Because $T_{\mathrm c}$ is extremely low at ambient pressure, detailed investigation of the superconducting state is experimentally challenging. The application of pressure enhances $T_{\mathrm c}$ to about 4~K at 2~GPa~\cite{2016YQ,Hu2019a,Takahashi2017,Piva2023,Heikes2018,Guguchia2017,hang2024}. This enhancement has motivated extensive studies on the possible realization of topological superconductivity. However, while the origin of the $T_{\mathrm c}$ enhancement remains unclear, the superconducting pairing symmetry is still under debate, with muon-spin rotation~\cite{Guguchia2017} and point-contact spectroscopy~\cite{hang2024} indicating possible $s^{++}$ or $s^{+-}$ states among several candidates.

From a microscopic perspective, it is desirable to employ experimental probes that can simultaneously access both the Weyl fermion states and the superconducting order. 
Nuclear magnetic resonance (NMR) is particularly suited for this purpose, as it can probe both the electronic states associated with superconductivity and those related to Weyl fermions. 

\begin{figure}[]
\centering
\includegraphics[width=0.5\linewidth]{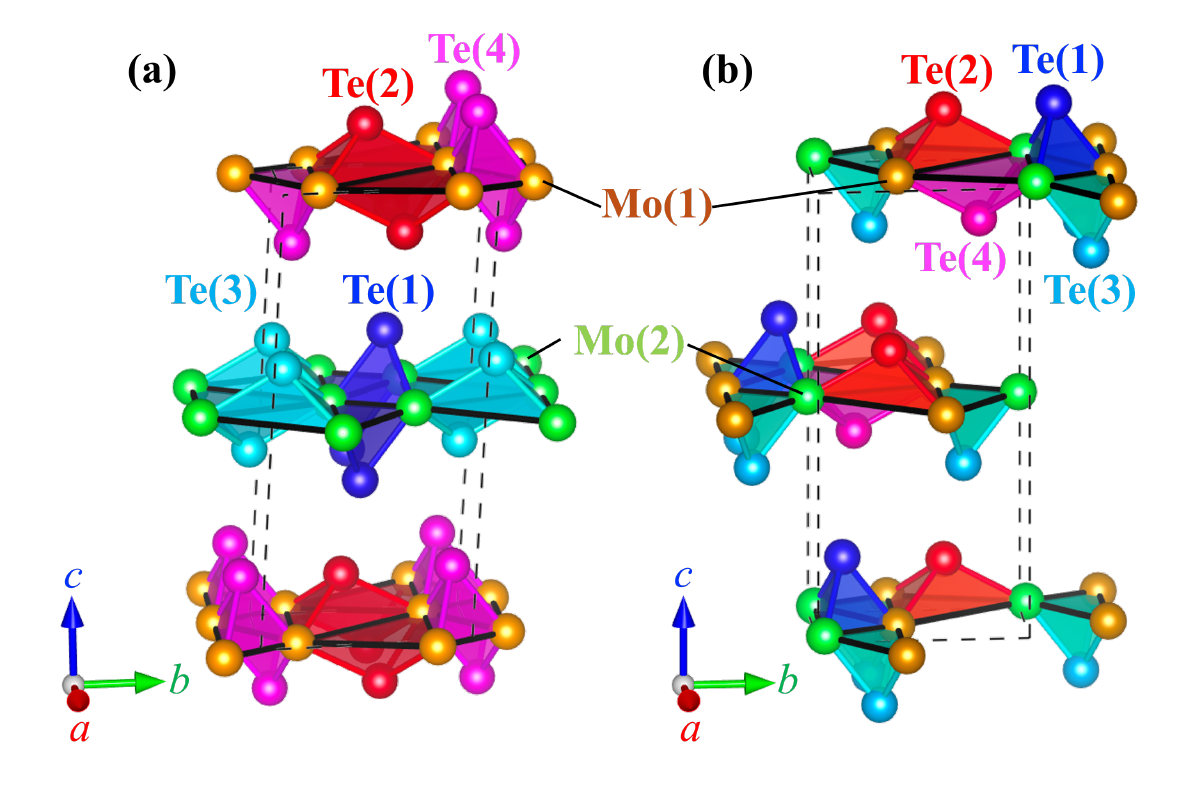}
\caption{\label{fig1}
(a) Crystal structure of the monoclinic 1T$^{\prime}$ phase. Each layer stacked along the c-axis contains only one type of Mo site, either Mo1 or Mo2. The cell parameters are $a=3.469 \AA$, $b=6.33 \AA$, $c=13.86 \AA$ and $\beta =93.92^{\circ}$ \cite{2016YQ}. (b) Crystal structure of the orthorhombic T$\mathrm{_d}$-MoTe$_2$. Layers are stacked along the $c$-axis, and both Mo1 and Mo2 sites coexist within the same layer. The cell parameters are $a=3.477 \AA$, $b=6.335 \AA$, $c=13.889 \AA$~\cite{2016YQ}.  Both the 1T$^{\prime}$ and T$\mathrm{_d}$ phases form a layered structure stabilized by van der Waals forces between the layers, with Mo and Te atoms occupying two and four inequivalent sites, respectively. Within each layer, distorted pyramidal units are formed, with three Mo atoms at the base and a Te atom at the apex. These pyramids are arranged in a zigzag pattern along the $a$-axis and can be broadly classified into two distinct types: in the 1T$^{\prime}$ phase, each layer contains a single type of Mo site, whereas the T$\mathrm{_d}$ phase features two inequivalent Mo sites that alternate along the $b$-axis. Additionally, the layers in the 1T$^{\prime}$ phase alternately contain different Te atom pairs (Te$(1,2)$ or Te$(3,4)$) along the $c$-axis, while the layers in the T$_{\rm d}$ phase all include four types of Te atoms. Despite these structural differences, a comparison of the local environments around the Te sites in both phases reveals only minor variations in the Mo-Te bond distances.}
\end{figure}

The main objective of this study is to investigate low-energy density of states (DOS) near the Fermi level under pressure and to clarify how they influence the pressure-induced enhancement of $T_{\mathrm c}$ in 1T-MoTe$_2$.
We measured the temperature and pressure dependence of the Knight shift ($K$) and the nuclear spin-lattice relaxation rate ($1/T_1$), which probe the static and dynamic spin susceptibilities, respectively, in both the 1T$^{\prime}$ and T$_{\mathrm d}$ phases, whose crystal structures are shown in Fig.~\ref{fig1}~(a) and (b).
The first-principles calculations of the DOS and the electric field gradient (EFG) were used to support the interpretation of the experimental results.
The calculated EFG values also allowed us to estimate the expected NQR frequencies, facilitating the identification of the observed NQR signals.
Section~\ref{sec4} outlines the experimental methods, and Section~\ref{sec5} presents and discusses the results, with conclusions summarized in Section~\ref{sec6}.

\section{Calculation of density of states (DOS) and electric field gradient (EFG)} \label{sec3} 

 \begin{figure}[b]
 \centering
\includegraphics[width=0.5\linewidth]{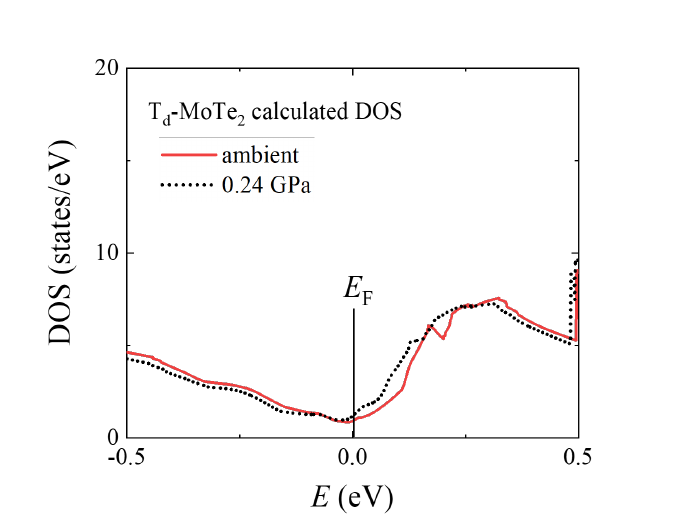}
\caption{\label{fig3-1}
Calculated total DOS as a function of energy in the T$\mathrm{_d}$ phase of MoTe$_2$. The red and black curves are the calculated DOS corresponding to ambient and 0.24 GPa, respectively.}
\end{figure}

The DOS calculation in the $T_{\rm d}$ phase was performed for ambient pressure and 0.24~GPa by the Density Functional Theory (DFT) with the full-potential local-orbital code (FPLO) ver.~22, employing a local-orbital basis set~\cite{Koepernik1999}. To determine the atomic position parameters used in the DOS calculation, we performed crystal structural optimization based on the relaxation of atomic positions using LDA calculations. This calculation was carried out using the WIEN2k package~\cite{Tran2009}, with the lattice parameters adopted from those reported in Ref. ~\cite{2016YQ}. For the calculation of the DOS using FPLO, we employed the LDA parametrized by Perdew and Wang\cite{Perdew1992} for the exchange-correlation potential, and the corresponding Brillouin zone was sampled using an $8\times8\times8$ k-mesh. The strong spin-orbit coupling was taken into account by performing full-relativistic calculations, wherein the Dirac Hamiltonian with a general potential is solved. 
The results of these calculations are presented in Fig. \ref{fig3-1}.

 The Mo nuclear quadrupole frequency, $\nu_Q$, can be obtained from the calculated EFG tensor at the Mo nuclear site. The EFG tensor elements, $V_{ij}$, are defined as the second partial derivative of the electrostatic potential $V(\boldsymbol{r})$ at the position of the nucleus: $V_{ij}=\left( \partial_i \partial_j V(0) - \frac{1}{3} \delta_{ij} \Delta V(0) \right)$, where $\delta_{i,j}$ is the Kronecker delta, with $\boldsymbol{r}$ being the position vector originating from the nucleus. The calculated $\nu_Q$ and the asymmetric parameter of EFG, $\eta$, for the two inequivalent Mo sites, Mo(1) and Mo(2), in the $\mathrm{T_d}$ phase are tabulated in Table~\ref{table:efg}. We note here that, according to the calculation, the distinct local environments of the two Mo sites in the ${\rm T_d}$ phase, as described in Fig.~\ref{fig1}, do not lead to significant differences both in $\nu_Q$ and $\eta$.
 \begin{table}[htb]
\caption{Calculated EFG parameters, $\nu_Q$ and $\eta$, for the Mo(1) and Mo(2) site in the ${\rm T_d}$ phase.
Expected two $^{97}$Mo resonance frequencies, $f_1$ and $f_2$, are also shown. See text for details.}
\label{table:efg}
\centering
\begin{tabular}{lcccc}\\
\hline
Phase & 
\begin{tabular}{@{}c@{}}$^{97}\nu_Q$\\(MHz)\end{tabular} & 
$\eta$ & 
\begin{tabular}{@{}c@{}}$f_1$\\(MHz)\end{tabular} & 
\begin{tabular}{@{}c@{}}$f_2$\\(MHz)\end{tabular} \\
\hline
Mo(1) & 5.84 & 0.87 & 9.42 & 10.52 \\
Mo(2) & 5.61 & 0.88 & 9.15 & 10.08 \\
\hline\end{tabular}
\end{table}

 \section{Experimental details} \label{sec4} 
 \subsection{Sample Preparation: powder and single crystal.}\label{subsec4.1} 
 The polycrystalline 1T-MoTe$_2$ was synthesized in a quartz ampoule sealed under vacuum, starting from a stoichiometric, intensively homogenized, mixture of element powders (Mo 99.95\% Alfa Aesar and Te 99.999\% Alfa Aesar) with the addition of 5~mg iodine as a reaction mediator. The reaction was carried out for 5~days at 600~$^\circ\mathrm{C}$ and a further 5 days at 1000 $^\circ\mathrm{C}$. After homogenization of the reaction product obtained, this thermal treatment was repeated a second time.
1T-MoTe$_2$ single crystals were grown from Te flux \cite{Wang2016,Yang2017}. Mo and Te raw elements were placed in an evacuated quartz tube in alumina crucibles in a ratio 1:25. The quartz tube was heated to 1000~$^\circ\mathrm{C}$ for two hours, then cooled to 820~$^\circ\mathrm{C}$ where excess Te flux was decanted in a centrifuge.  
Both samples synthesized in these ways were confirmed by x-ray diffraction to be in the 1T$^{\prime}$ phase at room temperature.

  \subsection{NMR/NQR experiments.}\label{subsec4.2} 
 Mo NMR measurements would be the most direct method to investigate the magnetism and superconductivity of 1T-MoTe$_2$. However, due to the weak signal intensities, we restricted our $^{97}$Mo NQR and $^{95, 97}$Mo NMR measurements to the spectrum of a powder sample at 4.2~K and ambient pressure. The NQR spectral measurements were performed by sweeping frequency, guided by the results of the DFT calculation (see Table~\ref{table:efg}). We succeeded in observing resonances corresponding to the two transitions for a nuclear spin $^{97}I = 5/2$, as described in Sec.~\ref{subsec5.2}. $^{95,97}$NMR spectral measurements were carried out by sweeping external magnetic field. Both the $^{95}$Mo and $^{97}$Mo nuclei have $I=5/2$. Obtained spectral data were analyzed using the gyromagnetic ratios $^{95}\gamma = 2.77$~MHz/T and $^{97}\gamma = 2.83$~MHz/T for the $^{95}$Mo and $^{97}$Mo, respectively.
 
 $^{125}$Te NMR measurements ($I=1/2$ and $^{125}\gamma = 13.45$~MHz/T) were performed under ambient pressure and high pressures up to 2.17~GPa.
The absence of nuclear quadrupole interactions simplifies the analyses of NMR spectrum and nuclear relaxation data. $^{125}$Te-NMR spectra were obtained by fast Fourier transformation (FFT) of the spin-echo signal. A uniaxial rotation system was employed for measuring the angular dependence of the spectrum on the magnetic field. Nuclear spin-lattice relaxation rate ($1/T_1$) was measured at the same peak frequency using the inversion recovery method. The $1/T_1$ values were extracted by fitting the recovery of the nuclear magnetization to the function $M(t)=M(\infty)(1-c_0\exp(-t/T_1))$, where $M(\infty)$ is the equilibrium nuclear magnetization, $c_0$ is a parameter indicating the degree of inversion ($c_0$=2 for complete inversion), and $t$ is the time after the inversion pulse.

  \begin{figure}[]
  \centering
\includegraphics[width=0.45\linewidth]{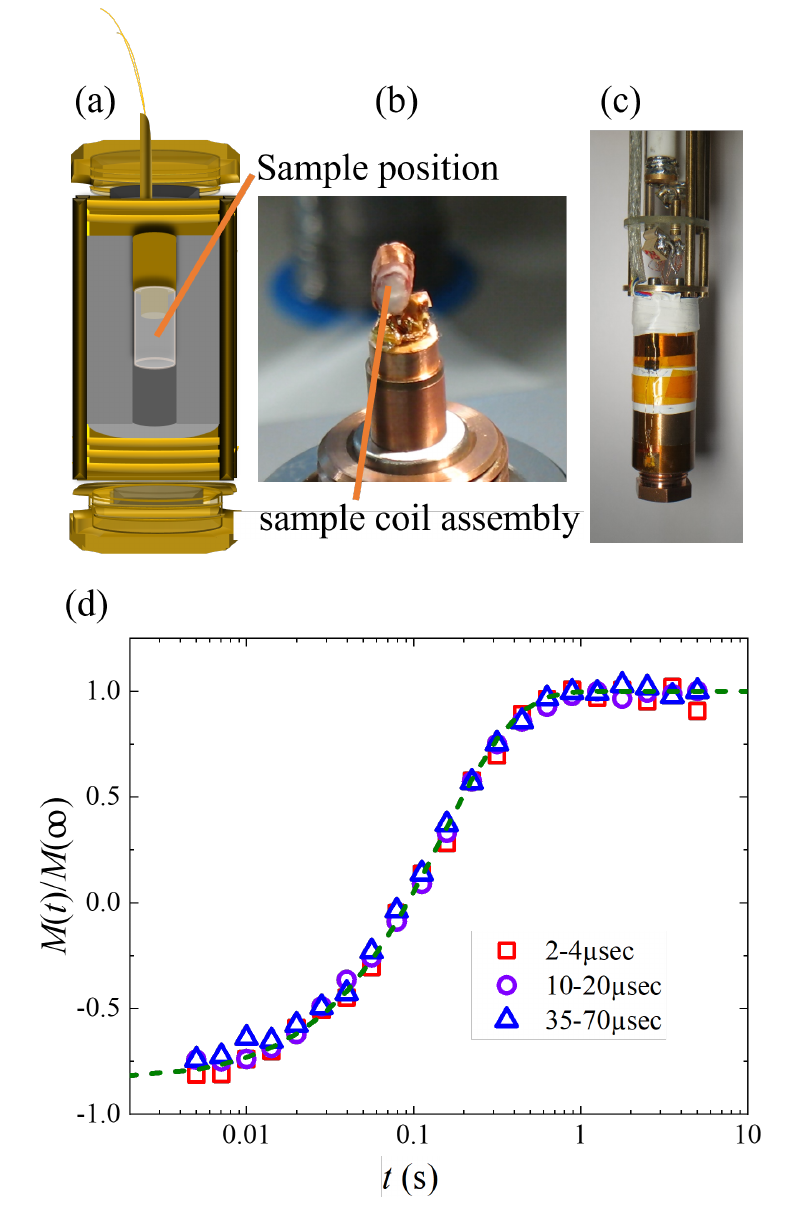}
\caption{\label{fig4.2-1}
(a) Schematic diagram of the self-clamped piston-cylinder cell. (b) NMR coil assembly. (c) NMR probe with the pressure cell and thermometers. (d)RF power dependence of nuclear magnetization recovery, $M(t)/M(\infty)$ vs $t$. The data were acquired at 4.2~K and 1.35~GPa, using an NMR measuring frequency of 6.235~MHz, and with the following sets of $\pi/2$ - $\pi$ pulse widths: 2 - 4 $\mu$sec, 10 - 20 $\mu$sec, and 35 - 70 $\mu$sec. The dashed line shows the theoretical curve with $T_1$=0.15~s. 
}
\end{figure}

 High-pressure measurements of $^{125}$Te-NMR were carried out using a self-clamped BeCu/NiCrAl piston-cylinder cell. The 1T-MoTe$_2$ powder sample, NMR coil, a manganin wire gauge, and a lead manometer were placed inside a Teflon capsule filled with pressure-transmitting medium, as shown in Figs.~\ref{fig4.2-1}(a), (b), and (c). A silicon-based organic liquid was used as the pressure medium. Pressure was determined by measuring the $T_{\mathrm{c}}$ of the lead. To ensure the absence of RF heating during the NMR experiment inside the cell, we examined the RF power dependence of the spin-echo shape and the recovery of nuclear magnetization after saturation. This was done by varying the excitation RF pulse width from 2 $\mathrm{\mu}$sec to 35 $\mathrm{\mu}$sec while maintaining the pulse inversion condition. As shown in Fig.~\ref{fig4.2-1}(d), the negligible difference in the recovery curves indicates the absence of RF heating under our experimental conditions.

  \section{Experimental results and discussions} \label{sec5} 
 \subsection{AC-susceptibility measurements.}\label{subsec5.1} 
 The macroscopic superconducting properties---particularly the upper critical field, $H_{\mathrm{c2}}$---have been investigated using AC susceptibility measurements. It is known that the onset of superconductivity causes a significant change in the impedance of a tuned RF circuit.

\begin{figure}[]
\centering
\includegraphics[width=0.5\linewidth]{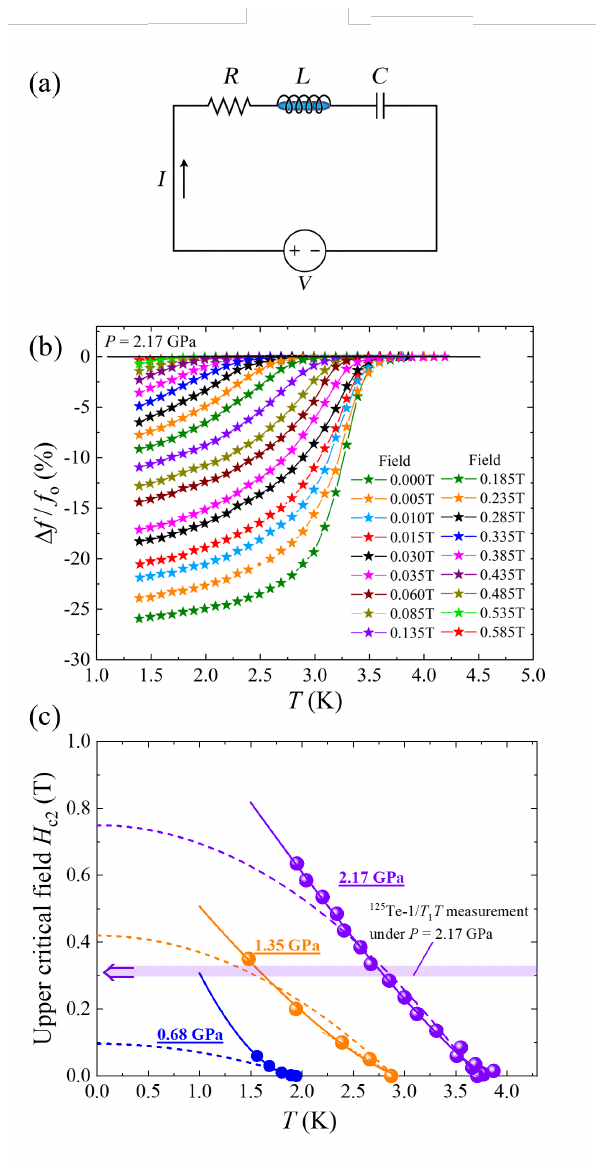}
\caption{\label{fig5-1}
(a) Equivalent circuit of NMR and ac susceptibility measurements. (b) Imaginary part of impedance of tune circuit measured by frequency cage, $(\Delta f)/f_0$= $(f_s-f_0)/f_0$ as a function of temperature under 2.17 GPa. (c) Temperature dependence of upper critical field, $H_{\mathrm{c2}}$, under 2.17 GPa (violet), 1.35 GPa (orange) and 0.68 GPa. The dashed lines are expected $H_{\mathrm{c2}}(T)$ from WHH model, and solid lines are fit to $H_{\mathrm{c2}}(T) = H_{\mathrm{c2}}(0)\times [1-T/T_c]^{\alpha}$ (see main text). }
\end{figure}

Figure~\ref{fig5-1}(a) shows a schematic diagram of a series RLC resonant circuit used for NMR measurements. The total impedance $Z_S$ of the circuit is expressed as $Z_S = R_S + iX_S$, where $R_S$ represents the effective resistance accounting for energy losses related to the loaded and unloaded quality factors ($Q_S$ and $Q_0$), as given by $R_S = G_1 \left( \frac{1}{2Q_S} - \frac{1}{2Q_0} \right)$. The reactance $X_S$ describes the frequency-dependent deviation from its resonant frequency $f_0$, which corresponds to the resonant frequency at 4.2~K, and is expressed as $X_S = G_2 \left( -\frac{f_S - f_0}{f} \right) + C.$ Here, $G_1$ and $G_2$ are scaling factors determined by circuit parameters, $f_S$ denotes the resonant frequency at each temperature and $C$ represents an offset constant. In this setup, the imaginary part of the impedance is directly related to superconducting diamagnetism via shifts in the resonant frequency. This provides an effective method for detecting the onset of superconductivity using the same coil employed for NMR measurements.

 As a representative example at 2.17~GPa, the fractional change in resonant frequency, $\Delta f/f_0$=$(f_s-f_0)/f_0$, is plotted as a function of temperature under various applied magnetic fields in Fig.~\ref{fig5-1}(b). The $T_{\mathrm{c}}$ were determined as the temperatures at which the resonant frequency dropped by 0.5\% below the normal-state value for each field.

Simple BCS theory predicts that the Cooper pair breaking depends on the combined effects of Pauli paramagnetism and orbital diamagnetism. If we follow the Werthamer-Helfand-Hohenberg (WHH) theory in the weak-coupling regime \cite{Werthamer1966}, the orbital-limiting upper critical field $H_{\mathrm{c2}}$ can be expressed as $H_{\mathrm{c2}}(T) = H_{\mathrm{c2}}(0) \left[ 1 - \left( \frac{T}{T_c} \right)^2 \right]$, where $H_{\mathrm{c2}}(0) = -0.693\,T_c \left( \frac{dH_{\mathrm{c2}}}{dT} \right)_{T_c}$ and $\left( \frac{dH_{\mathrm{c2}}}{dT} \right)_{T_c}$ is the initial slope of $H_{\mathrm{c2}}(T)$ at $T_{\mathrm{c}}$. Dashed lines in Fig.~\ref{fig5-1}(c) show these fits for 2.17, 1.35, and 0.68~GPa. Although we obtained zero-temperature upper critical fields of $H_{\mathrm{c2}}(0) = 0.75$~T, $0.42$~T, and $0.096$~T for 2.17~GPa, 1.35~GPa, and 0.68~GPa, respectively, a clear deviation from the experimental data is observed at lower temperatures. To address this discrepancy, we used the phenomenological expression
$H_{\mathrm{c2}}(T) = H_{\mathrm{c2}}(0) \left[ 1 - \left( \frac{T}{T_c} \right) \right]^\alpha$
to fit the data~\cite{Muller2001}. 
We obtained $H_{\mathrm{c2}}(0)$, $T_{\mathrm{c}}$, and $\alpha$ for 2.17~GPa, 1.35~GPa, and 0.68~GPa, which are summarized in Table~\ref{table2}.

The increase of $H_{\mathrm{c2}}(0)$ with pressure and the fact that $\alpha$ is greater than one suggest that 1T-MoTe$_2$ belongs to a strong-coupling regime under high pressures.

 \begin{table}[htbp]
\centering
\caption{Pressure dependence of $T_{\mathrm{c}}$, $H_{\mathrm{c2}}(0)$, and $\alpha$ obtained from fitting.}
\label{table2}
\begin{tabular}{cccc}
\hline
\textbf{Pressure (GPa)} & \boldmath$T_{\mathrm{c}}$ \textbf{(K)} & \boldmath$H_{\mathrm{c2}}(0)$ \textbf{(T)} & \boldmath$\alpha$ \\
\hline
0.68 & 1.94 & 1.14 & 1.8 \\
1.35 & 2.97 & 1.21 & 1.4 \\
2.17 & 3.81 & 1.50 & 1.1 \\
\hline
\end{tabular}
\end{table}

 \subsection{Mo NQR and NMR at ambient pressure}\label{subsec5.2} 
 The $^{97}$Mo NQR spectrum at 4.2~K in the T$_{\mathrm d}$ phase is shown in Fig.~5(a), which was successfully identified with the help of the DFT calculation (see Table~I). The observed peak frequencies $f_1$ and $f_2$ are assigned to the transitions between the nuclear quadrupolar split energy levels of $3/2 \leftrightarrow 1/2$ and $5/2 \leftrightarrow 3/2$, respectively.

 
 From these values, $\nu_Q$ and $\eta$ are determined to be $^{97}\nu_Q = 0.54$ and $\eta = 0.73$, respectively. For $f_1$ and $f_2$, the maximum discrepancy between the prediction by the DFT calculation and the experimental value is approximately $16\%$, demonstrating that highly accurate DFT calculations prove to be a powerful tool in the detection of unknown NQR signals. It should be noted that because of the small differences in $\nu_Q$ and $\eta$ between  the Mo(1) and Mo(2) sites, as suggested by the DFT calculation, the site differentiation was not observed in the present NQR spectra. 

\begin{figure}[]
 \centering
\includegraphics[width=0.5\linewidth]{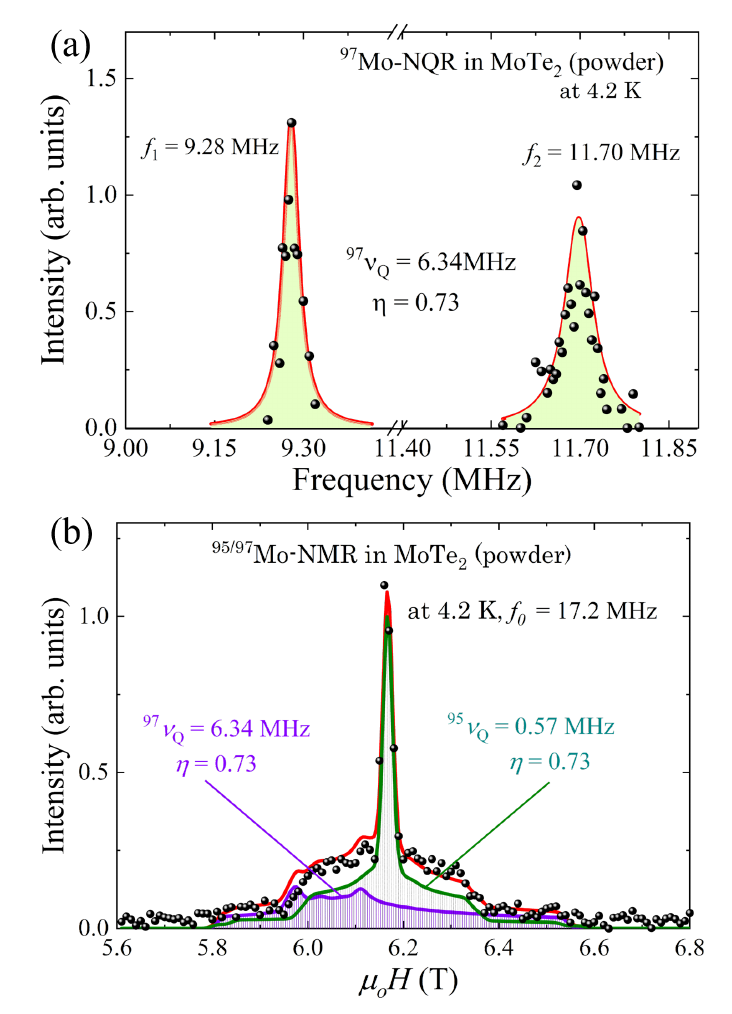}
\caption{\label{fig5-2}
(a) Frequency-swept $^{97}$Mo NQR spectrum measured at 4.2 K. From the two observed resonances at $f_1 = 9.28$~MHz and $f_2 = 11.70$~MHz, we obtained $^{97}\nu_Q = 6.34$~MHz and $\eta = 0.73$.  Field-swept $^{95}$Mo and $^{97}$Mo NMR spectra measured using a measuring frequency of 17.2~MHz at 4.2~K. The green and purple lines are simulated spectra using $^{95}\nu_Q = 0.57$~MHz for $^{95}$Mo and $^{97}\nu_Q = 6.34$~MHz for $^{97}$Mo, with $\eta = 0.73$ being common to both. Here, $^{95}\nu_Q$ was calculated from the result of $^{97}\nu_Q = 6.34$~MHz and the ratio of nuclear quadrupole moment $^{95}Q/^{97}Q = 0.086$. The red line is the sum of them.
}
\end{figure}

Field-swept $^{95/97}$Mo-NMR powder spectra obtained at 4.2~K are shown in Fig.~\ref{fig5-2}(b). The observed spectrum is composed of $^{95}$Mo and $^{97}$Mo NMR signals with a typical powder pattern. Due to the one order of magnitude larger quadrupole moment of $^{97}$Mo, we expect to observe only a second-order powder spectrum for the central transition. Using the same EFG parameters obtained from the NQR spectra and taking into account the natural abundance ratio of $^{95}$Mo and $^{97}$Mo, we  simulated the NMR spectrum by convoluting $^{95}$Mo and $^{97}$Mo components, as shown in Fig.~\ref{fig5-2}(b). Here, the Mo(1) and Mo(2) sites were treated as being indistinguishable. The consistency between the simulated and observed spectra, along with the agreement between the theoretically predicted and experimental NQR spectra, clearly demonstrates the high reliability of the DFT calculation, thereby enabling subsequent discussion based on the DOS.

  \subsection{Te NMR under pressure: Spectra and Knight Sift}\label{subsec5.3} 
 
\begin{figure}[]
\centering
\includegraphics[width=0.5\linewidth]{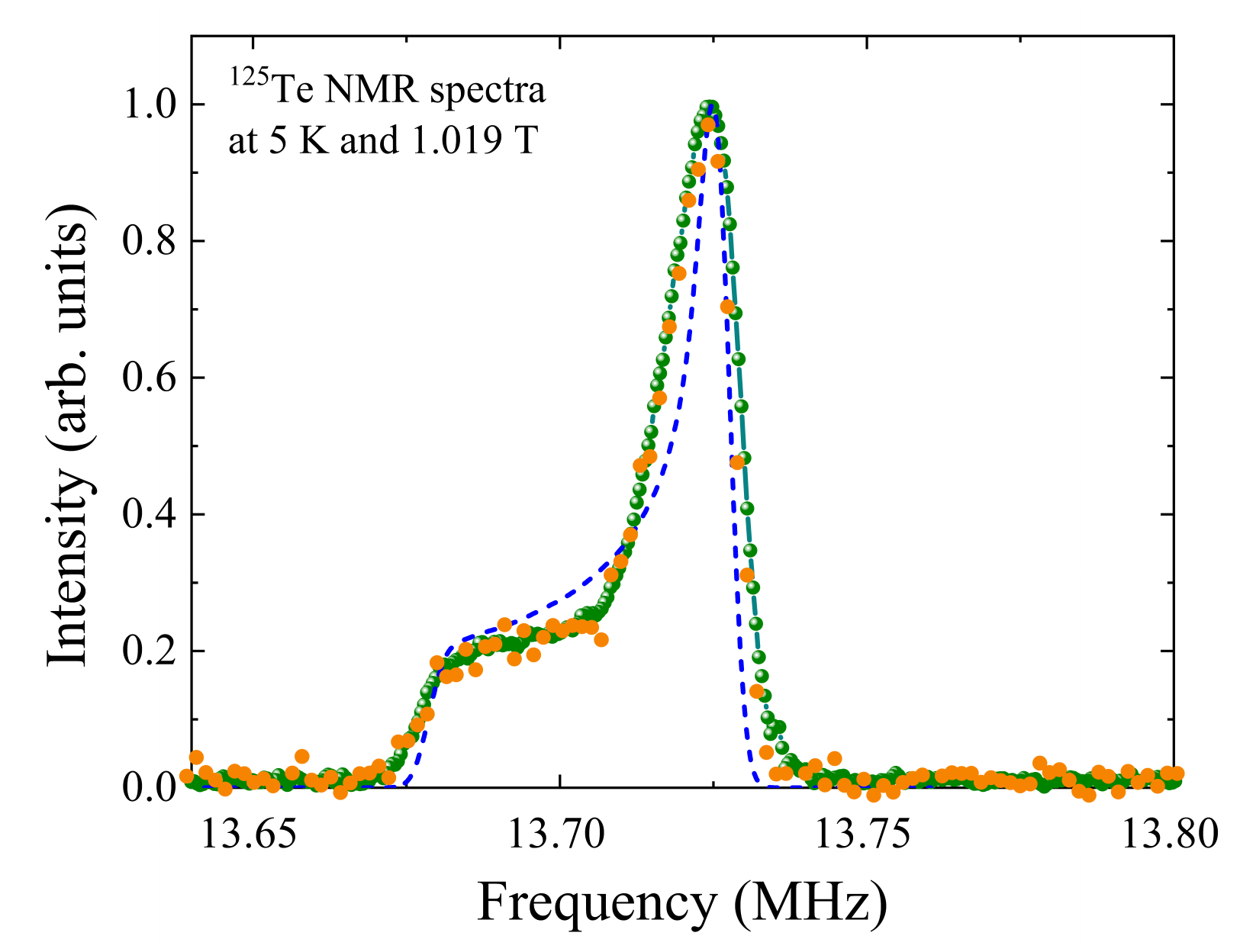}
\caption{\label{fig5-3-1}
A typical $^{125}$Te-NMR spectrum at 5 K and 1.019 T is shown. The frequency spectrum was obtained by performing an FFT on the spin echo signal (green dots connected by a line) and is compared with data acquired using the field sweep method (orange circles). The dashed line represents a simulated axial symmetry powder pattern with $K_{\mathrm {iso}}$ = 0.02\% and $K_{\mathrm {ax}} = -0.05\%$.}
\end{figure}

As mentioned in Sec.\ref{subsec4.2}, $^{125}$Te NMR is a highly suitable tool for investigating the microscopic static and dynamic properties of 1T-MoTe$_2$. However, data analysis can be challenging due to the frequent presence of unknown transferred hyperfine interactions at the Te sites. Despite this drawback, the nuclear spin $I = 1/2$ and relatively large gyromagnetic ratio ($\gamma_n = 13.45$~MHz/T) provide significant advantages: they enable signal detection under high-pressure and high-temperature conditions and facilitate data analysis without the complications arising from nuclear quadrupole interactions. As an example, the $^{125}$Te NMR spectrum is shown in Fig.~\ref{fig5-3-1}.

In general, these data can be obtained either by performing an FFT of the spin-echo signal at a constant magnetic field or by using the frequency (or field) sweep method. The former is more suitable for relatively narrow lines, while the latter is better suited for broad lines. We first measured the NMR spectrum at 5~K using both methods and compared the results. As shown in Fig.~\ref{fig5-3-1}, the spectra from both methods matched perfectly. Therefore, we primarily used the FFT method to acquire data for temperature and pressure dependence studies. It should be noted that the spectrum shown in Fig.~\ref{fig5-3-1} is a typical powder pattern with uniaxial symmetry. Under this assumption, we successfully obtained the best simulated spectrum with $K_\mathrm{iso} = 0.02\%$ and $K_\mathrm{ax} = -0.05\%$, as shown by the dashed line in Fig.~\ref{fig5-3-1}.

\begin{figure}[]
\centering
\includegraphics[width=0.5\linewidth]{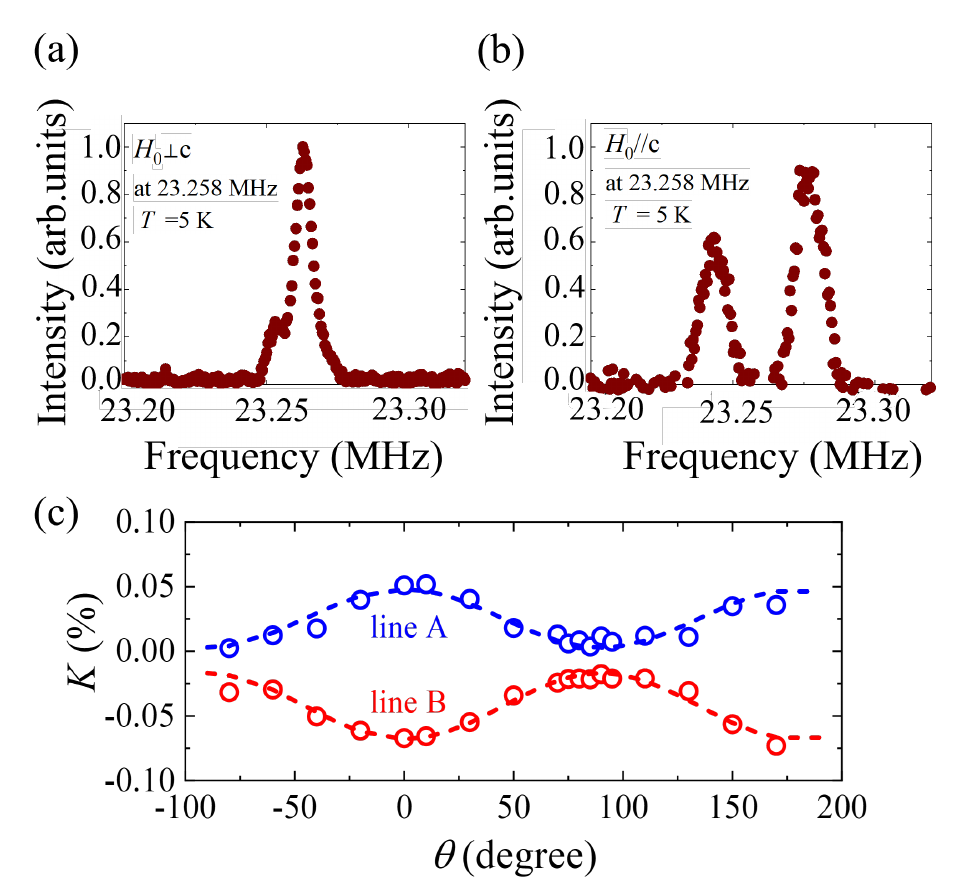}
\caption{\label{fig5-3-2}
Angular dependence of $^{125}$Te NMR spectra at 5 K and 23.258 MHz:(a) Applied magnetic field parallel to the $c$-axis A: (b) Applied field perpendicular to the c-axis B. (c) Angular dependence of the $K$ for two distinct groups of Te sites. The dashed lines represent the function $K(\theta)=K_{\mathrm iso}+K_{\mathrm ax}(3\cos^2\theta-1)$. These results are consistent with the 90$^{\circ}$ rotational symmetry of the orthorhombic crystal structure.}
\end{figure}

We also successfully observed $^{125}$Te NMR spectra in a single crystal. The $\theta$-dependence of the spectrum, where $\theta$ is the angle between the crystal $c$-axis and the magnetic field $H_0$, was measured using a uniaxial rotation system. As shown in Fig.~\ref{fig5-3-2}, two distinct groups of resonance lines (A and B) were observed, and both lines exhibit two-fold rotational symmetry with respect to $\theta$, reflecting the orthorhombic structure of the T$_{\mathrm d}$ phase. The two observed lines likely originate from two pairs of the four inequivalent Te sites, i.e. (Te(1), Te(3)) and (Te(2), Te(4)), that have a similar local environment surrounded by the Mo atoms. Thus, although the T$_{\mathrm {d}}$ phase lacks inversion symmetry along the $c$-axis, no phenomena stemming from this unique structure were observed in the present measurements. The $\theta$-dependence of the spectrum reveals that the lines A and B become proximate to each other, when $H_0 \perp c$. This behavior is consistent with the assumption of uniaxial symmetry used in the powder spectrum fit (see Fig.~\ref{fig5-3-1}). The $\theta$ dependence of the Knight shift, $K(\theta)$, can be fitted by a typical expression for an axially symmetric system: $K(\theta) = K_{\mathrm iso} + K_{\mathrm ax} (3\cos^2 \theta -1)$, yielding $K_{\mathrm {iso}}= 0.018 \%$ and $K_{\mathrm {iso}}= 0.015 \%$ for line A; and $K_{\mathrm {iso}}= -0.034 \%$ and $K_{\mathrm {ax}}= -0.017 \%$ for line B. However, due to the limitation of our uniaxial rotation system, we have not obtained the comprehensive line profile of the single-crystal necessary to examine the overall consistency with the powder pattern. Therefore, we will focus our discussion on the powder data in the following sections, and our future efforts will involve further detailed spectral measurements of the field angular dependence.

\begin{figure}[]
\centering
\includegraphics[width=0.6\linewidth]{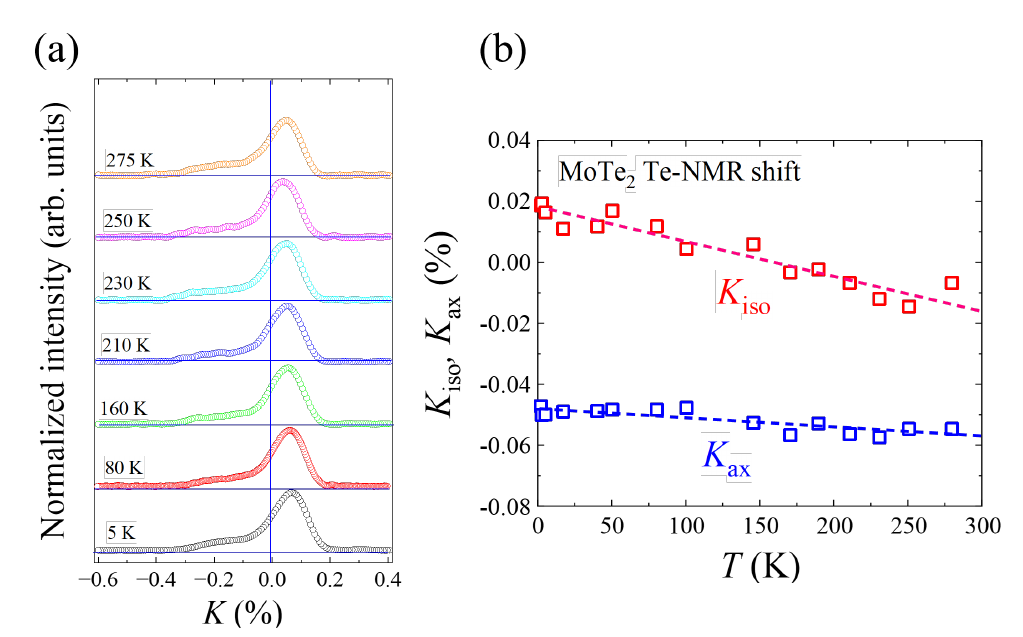}
\caption{\label{fig5-3-3}
(a) Temperature dependence of FFT spectrum measured at 1.0186~T. (b) Temperature dependence of $K_{\mathrm {iso}}$ and $K_{\mathrm {ax}}$ under ambient pressure.}
\end{figure}

\begin{figure}[]
\centering
\includegraphics[width=0.5\linewidth]{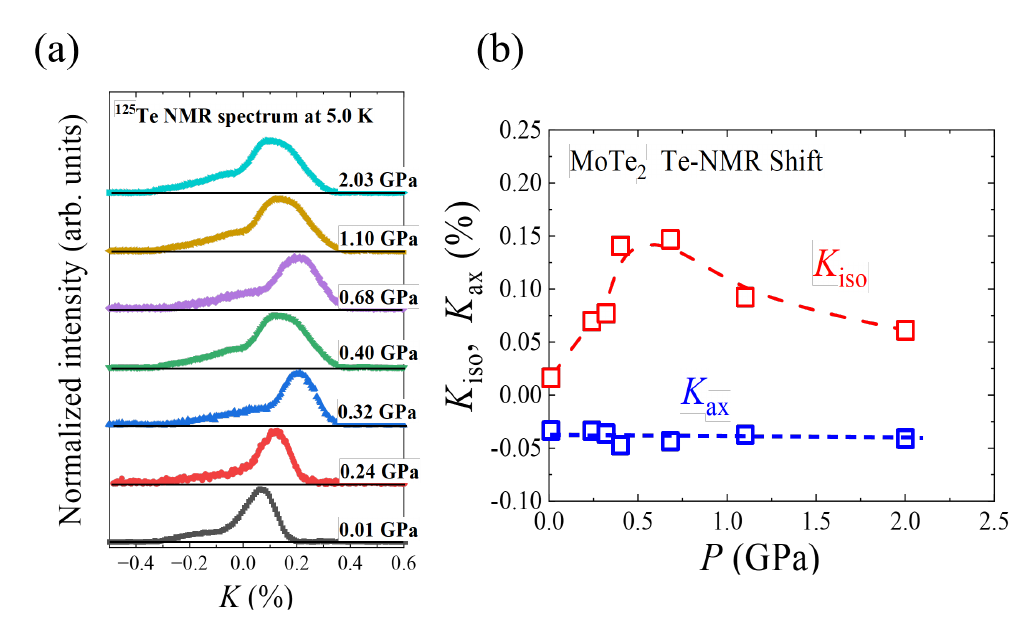}
\caption{\label{fig5-3-4}
(a) Pressure dependence of FFT spectrum. (b) Pressure dependence of $K_{\mathrm {iso}}$and $K_{\mathrm {ax}}$ measured at 5~K.}
\end{figure}

The temperature dependence of $^{125}$Te NMR powder spectrum at ambient pressure and its pressure dependence of at 5~K were measured. The results are shown in Figs.~\ref{fig5-3-3}(a) and \ref{fig5-3-4}(a), respectively. Also extracted $K_{\mathrm {iso}}$ and $K_{\mathrm {ax}}$ are displayed in Figs.~\ref{fig5-3-3}(b) and \ref{fig5-3-4}(b), respectively. $K_{\mathrm {ax}}$ shows little dependence on either temperature or pressure. In contrast, the $K_{\mathrm{iso}}$ gradually decreases with increasing temperature, suggesting the temperature-dependent magnetic susceptibility. Furthermore, $K_{\mathrm{ax}}$ exhibits a distinctive pressure dependence: it initially increases rapidly up to about 0.7 GPa, and since this behavior is consistent with the pressure dependence of $T_{\mathrm c}$, these behaviors are likely related to the DOS at the Fermi level. Upon further pressure increase, $K_{\mathrm {iso}}$ begins to decrease, and its overall pressure dependence closely resembles that of the nuclear spin-lattice relaxation rate divided by temperature, $1/T_1T$, which will be discussed in Section \ref{subsec5.4.1}.

It should also be noted that 1T-MoTe$_2$ undergoes a first-order phase transition from the high-temperature monoclinic 1T$^{\prime}$ phase to the low-temperature orthorhombic T$\mathrm{_d}$ phase at 250~K~\cite{2016YQ,Hu2019a,Takahashi2017,Piva2023,Heikes2018,Guguchia2017}. The T$\mathrm{_d}$ phase reverts to the 1T$^{\prime}$ phase under increasing pressure, with the 1T$^{\prime}$ phase being stabilized above 1~GPa at all temperatures~\cite{Piva2023,Heikes2018}. However, the Te-NMR spectra do not exhibit any signs of this phase transition, nor do they show any site distinction. This indicates that the transferred hyperfine interaction at the Te sites is insensitive to changes in crystal symmetry.

 \subsection{Te NMR under pressure: Nuclear spin-lattice relaxation}\label{subsec5.4} 
 \subsubsection{Normal state}\label{subsec5.4.1}

 We measured the $^{125}$Te NMR $T_1$ at the main peak position of the spectrum at pressures up to 2.17~GPa. 
 When the relaxation process is magnetic, which is the case for $^{125}$Te with $I = 1/2 $, $1/T_1T$ can be generally expressed by the dynamical susceptibility $\chi (\boldsymbol{q}, \omega)$, where $\boldsymbol{q}$ is the wave vector and $\omega$ is the frequency, as shown in the following equation~\cite{Narath1968,Moriya1963}:
 \begin{equation}
\left( \frac{1}{T_1T} \right) = \frac{2\gamma_N^2 k_\mathrm B}{g^2 \mu_\mathrm B^2} \sum_{\mathbf{q}} {A_\mathbf{q}^{2}} \frac{\chi''_{\perp}(\mathbf{q}, \omega)}{\omega_N}.
\label{eq:5.3-1}
\end{equation}
Here, $\gamma_N$ is the gyromagnetic ratio, $k_\mathrm B$ is a Boltzmann constant, $g$ is called the $g$-factor, $\mu_\mathrm B$ is a Bohr magneton, $\chi''_{\perp}(\mathbf{q}, \omega)$ is the transverse component of the imaginary part of dynamical susceptibility, $\omega_N$ is the NMR frequency, $A_\mathbf q$ is the $\mathbf q$-dependent hyperfine coupling constant. In MoTe$_2$, the hyperfine field at the Te sites originates from both the Fermi contact interaction with conduction electrons and the transferred hyperfine interaction with neighboring Mo moments through $d$-$p$ mixing. Because our experimental observations provide clear evidence that the former mechanism dominates the relaxation process—at least at low temperatures—we analyze the experimental results using an electron-band framework within the random phase approximation (RPA) as follows:

\begin{equation}
\left( \frac{1}{T_1T} \right) \propto \frac{A_\mathrm{hf}^2}{T} \int f(E)[1 - f(E)] \{N(E)\}^2 dE,
\label{eq:5.3-2}
\end{equation}
where $A_\mathrm{hf}$ is the $\mathbf{q}$-averaged hyperfine coupling constant, $N(E)$ is the energy-dependent DOS near the Fermi level, and $f(E)$ is the Fermi distribution function.

\begin{figure}[]
\centering
\includegraphics[width=0.5\linewidth]{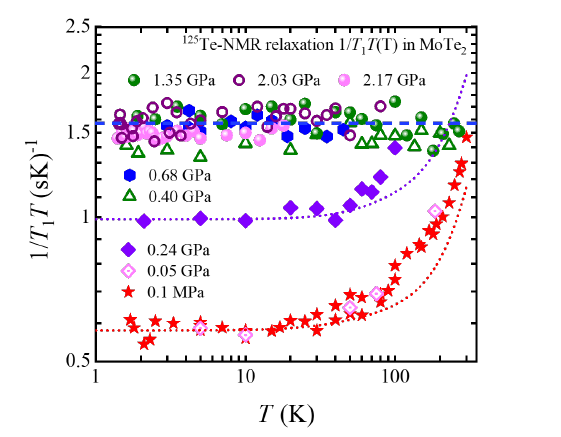}
\caption{\label{fig5-4-1}
Temperature dependence of $1/T_T$ measured at the Te site in the T$_{\mathrm d}$ phase up to 2.17~GPa.
The red and violet dashed lines represent calculated $1/T_T(T)$ at ambient pressure and 0.24~GPa, respectively, based on the DOS shown in Fig.~2.
}
\end{figure}

The temperature dependence of the $1/T_1T$ is shown in Fig.~\ref{fig5-4-1}. The data exhibit a Korringa-type behavior, characterized by temperature independence of $1/T_1T$ below approximately 30~K. In contrast, $1/T_1T$ increases rapidly above $T \sim 30$~K. According to our DFT calculation, the energy dependence of the DOS forms a V-shaped DOS around $E_\mathrm{F}$, as shown in Fig.~\ref{fig3-1}. This plausibly leads to the increase in $1/T_1T$ with increasing temperature due to the contribution of the DOS at energy levels away from $E_\mathrm F$. Another notable feature is the strong pressure dependence of the Korringa term at relatively low pressures: the temperature independent $1/T_1T$ increases (from 0.6) to $1.5 ({\mathrm sec} \cdot {\mathrm K})^{-1}$ with increasing the pressure to approximately 0.7~GPa.


Indeed, these trends are reproduced by computing Eq.~(2) using the DOS at ambient pressure and 0.24~GPa in the T$_{\mathrm d}$ phase shown in Fig.~\ref{fig3-1}. Here, since $A_{\mathrm {hf}}$ is unknown, the calculated values of $1/T_1T$ are normalized to the experimental value obtained at low temperature. The calculated $1/T_1T$ at ambient pressure, shown by the red dotted curve in Fig.~\ref{fig5-4-1}, increases at high temperatures, consistent with the experimental results, despite the calculated increase being slightly shifted toward higher temperatures compared to the measured temperature dependence. Moreover the constant $1/T_1T$ value at low temperatures increases at 0.24~GPa, while the similar temperature dependence is maintained, as shown by the violet dotted curves in Fig.~\ref{fig5-4-1}. These trends are in overall good agreement with the experimental results.

 \begin{figure}[]
 \centering
\includegraphics[width=0.4\linewidth]{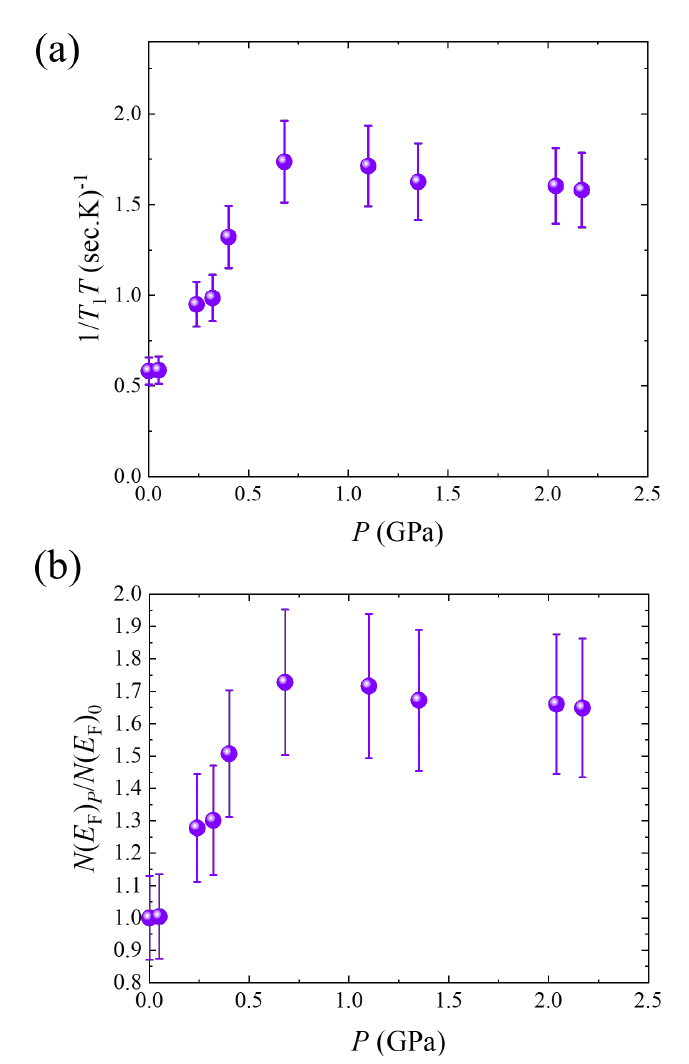}
\caption{\label{fig5-4-2-1}
(a) Pressure dependence of the low-temperature $1/T_T$. Pressure dependence of normalized DOS $N_s(E_F)_p / N_s(E_F)_0$ estimated using the Korringa relation described by Eq.~(3).
}
\end{figure}
 
Next, we discuss the pressure dependence of the low-temperature $1/T_1T$ and $T_{\mathrm c}$ based on the simplified expression for the Korringa rate and the BCS expression for $T_{\mathrm c}$. We here adopt the equation of the Korringa relation for weakly interacting electron systems ~\cite{Slichter1990,Korringa1950P} described next: 
\begin{equation}
\left( \frac{1}{T_1T} \right)_s = \frac{4\pi k_B}{\hbar} (\gamma_n \hbar H_\mathrm{hf}^s)^2 \left( N(E_{\rm{F}}) \right)^2,
\label{eq:5.3-3}
\end{equation}
where $\left( \frac{1}{T_1T} \right)_s$ represents the spin component of $1/T_1T$ and $H_\mathrm{hf}^s$ is the hyperfine field per spin at the nuclear site. The pressure dependences of the constant $1/T_1T$ at low temperature and $N(E_{\mathrm F})_P/N(E_{\mathrm F})_{0}$ estimated from the relation in Eq.~\ref{eq:5.3-3} are shown in Fig.\ref{fig5-4-2-1}(b), where $N(E_\mathrm F)_P$ is normalized to unity at ambient pressure.  It should be noted that the DOS increases almost linearly with pressure up to about 0.7~GPa, followed by a slightly decrease at higher pressures. On the other hand, $T_{\mathrm c}$ is described by the following well-known BCS formula~\cite{Bardeen1957}~: 

\begin{equation}
T_\mathrm c = 1.13\Theta_D \exp \left[-\frac{1}{V_0 N(E_\mathrm F)}\right],
\label{eq:5.3-5}
\end{equation}
where $\Theta_D$ is the Debye temperature and $V_0$ is the interaction strength.

\begin{figure}[]
  \centering
\includegraphics[width=0.5\linewidth]{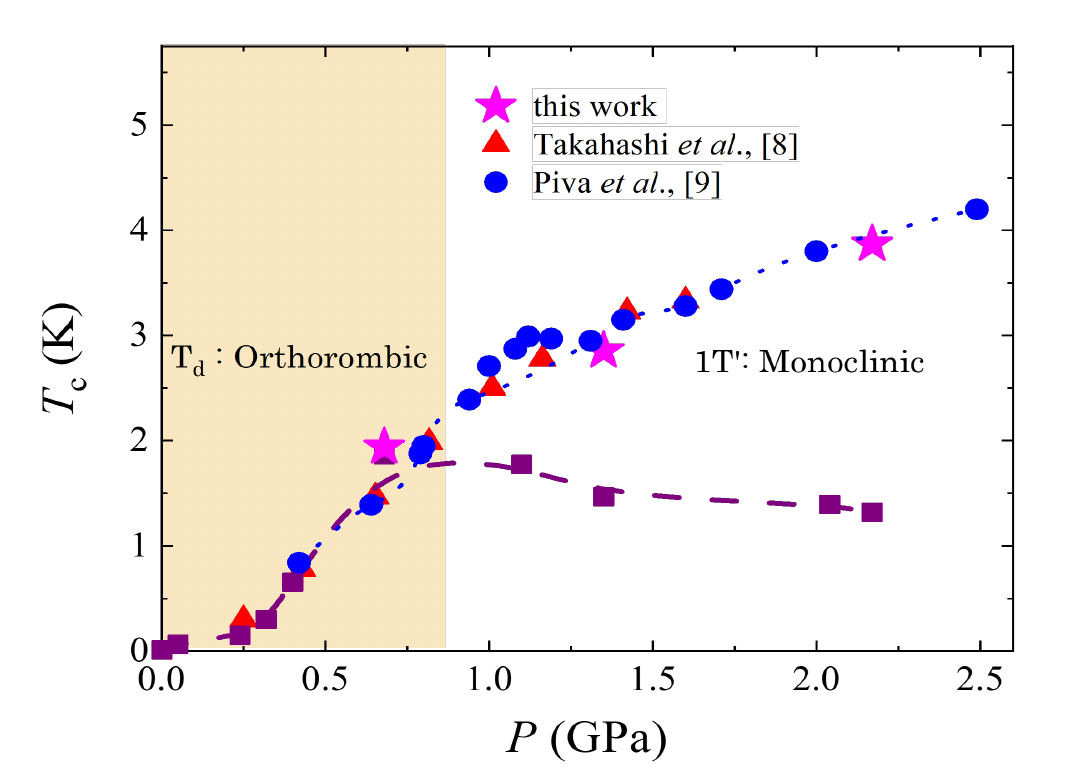}
\caption{\label{fig5-4-2-2}
Comparison of the pressure dependence of $T_\mathrm c$ (red and blur circles) from Refs.~8 and 9, respectively, and that expected from the BCS relation between $T_\mathrm c$ and the DOS (Fig.~11). See text for details. The latter is normalized to match the measured $T_\mathrm c$ at ambient pressure.}
\end{figure}

If we assume no significant change of $\Theta_D$ and $V_0$ with pressure, we can simply compare the pressure dependences of $T_{\mathrm c}$ and $N(E_\mathrm F)_P/ N(E_\mathrm F)_{0}$ (shown in Fig.~11) via the relation $T_c(P) = T_c(0) \exp \left[ -a \left\{ \left( \frac{N(E_\mathrm F)_P}{ N(E_\mathrm F)_{0}} \right)^{-1} -1 \right\} \right]$, where $a$ is a constant. Such plots are shown in Fig.~\ref{fig5-4-2-2}. We can immediately observe that below $\sim$0.7~GPa (in the $\mathrm{T_d}$ orthorhombic phase), the pressure dependence of $T_{\mathrm c}$ determined experimentally as shown in Fig.~\ref{fig5-4-2-2} is in general agreement with that of $N(E_\mathrm F)$. However, above $\sim$0.7~GPa, the experimental $T_{\mathrm{c}}$ exhibits a distinctly different behavior from what is expected from the pressure dependence of $N(E_\mathrm F)$ shown in Fig.~\ref{fig5-4-2-1}(b). This suggests the presence of an additional pairing mechanism beyond the conventional phonon-mediated BCS framework—possibly involving spin fluctuations that are enhanced $T_{\mathrm{c}}$ under pressure.

 \subsubsection{Superconducting state}\label{subsec5.4.2} 

\begin{figure}[]
   \centering
\includegraphics[width=0.5\linewidth]{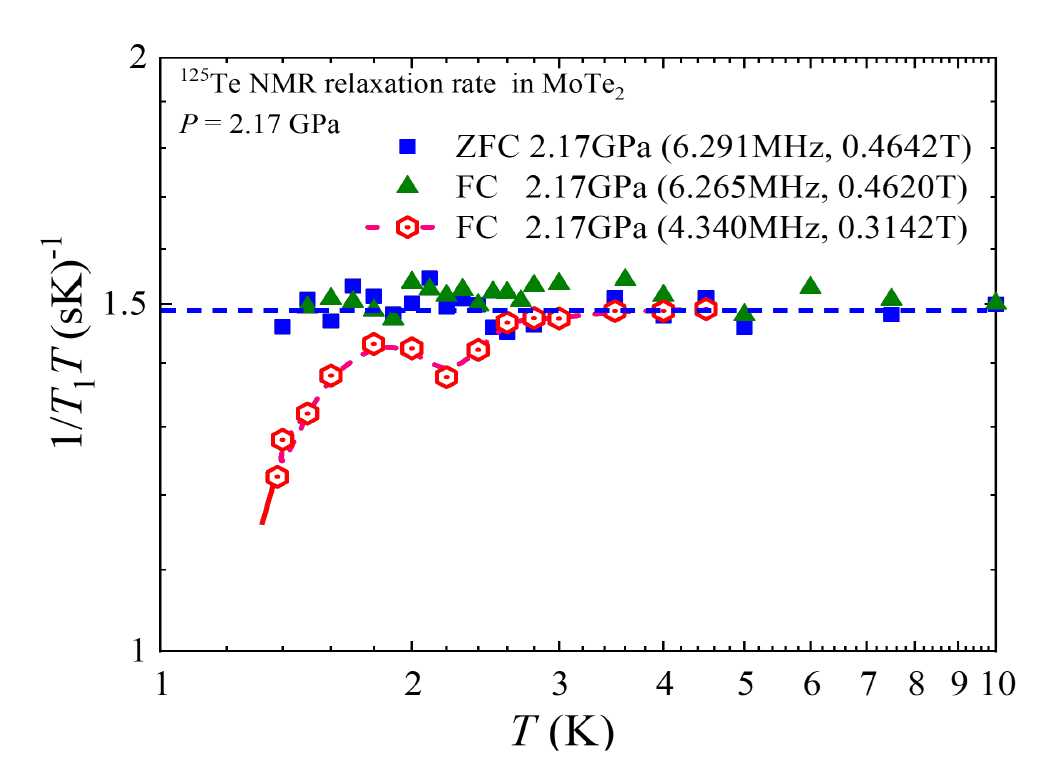}
\caption{\label{fig5-4-2}
Temperature dependence $1/T_1T$ measured by $^{125}$Te NMR after three different cooling procedures. Blue and green squares represent the data measured at 6.291~MHz after zero-field cooling (ZFC) and at 6.265~MHz after field cooling (FC), respectively. Dotted red circles represent the data measured at 4.340~MHz after field cooling.}
\end{figure}

 In order to gain deeper insights into the mechanism of superconductivity in 1T-MoTe$_2$, it is essential to clarify the order parameter below $T_{\mathrm{c}}$. Typically, this will be achieved by studying the temperature and pressure dependence of the nuclear relaxation time, $T_1$. Unfortunately, the current NMR equipment is unable to access the low temperatures and high pressures required for such measurements. Despite this limitation, we attempted to measure the temperature dependence of $T_1$ at our highest pressure (2.17~GPa), using the lowest NMR frequency (4.34~MHz, corresponding to an applied field of 0.314~T) and down to the lowest temperature of 1.45~K.
We present the results in Fig.~\ref{fig5-4-2}, where the measurements were carried out after three different cooling procedures. Initially, we measured $1/T_1T$ under zero-field-cool and field-cool conditions at 6.29~MHz (0.464~T). As shown by blue and green squares, no evidence for the onset of superconductivity was observed. However, when we reduced the frequency to 4.34~MHz corresponding to 0.31~T, a signature of the onset of superconductivity was observed (see the red dotted circles in Fig.~\ref{fig5-4-2}). As indicated by Fig.~\ref{fig4.2-1} (c), the AC-susceptibility measurement of the present sample predict $T_{\mathrm{c}}$ should be around 2.7~K, which just corresponds to the point where the reduction of $1/T_1T$ begins with cooling. Although we need more data across different fields, the following points can be argued:
1) There is no increase in $1/T_1T$ just above $T_{\mathrm{c}}$ (the so called coherence peak) characteristic to the simple phonon mediated $s$-wave superconductivity, 2) The temperature dependence of $1/T_1T$ exhibits a characteristic two-step decrease which cannot be expressed by the single superconducting gap, indicating that 1T-MoTe$_2$ has a multi-gap superconducting state. This has been argued by $\mathrm{\mu}$SR measurements at pressure between 0.45~GPa and 1.3~GPa~\cite{Guguchia2017}.
The present data does not give us a conclusive result for the superconductivity associated with the topologically trivial 1T$^{\prime}$-phase of MoTe$_2$. Nevertheless, there exist several important hints which motivate further experiments to explore broader pressure and temperature ranges to refine the data and gain more conclusive evidence. 

 \section{Concluding remarks} \label{sec6} 

 We have employed AC-susceptibility and microscopic techniques—specifically Nuclear Magnetic Resonance (NMR) and Nuclear Quadrupole Resonance (NQR)—to investigate the fundamental properties of the pressure-induced topological superconductivity candidate 1T-MoTe$_2$. Our study includes detailed measurements of the pressure dependence of superconducting transition temperature $T_{\mathrm c}$ and upper critical field ($H_{\mathrm{c2}}$), and the temperature dependence of NMR/NQR line shapes, Knight shift, and nuclear spin-lattice relaxation rate ($1/T_1$). 

\begin{itemize}
    \item The $T_\mathrm c$ was determined from the resonance frequency shift of the NMR tuning circuit and its magnetic field and pressure dependences were investigated. By fitting the field dependence of $T_\mathrm c$ using the phenomenological relation $H_{\mathrm{c2}}(T) = H_{\mathrm{c2}}(0) \times [1 - (T/T_{\mathrm{c}})]^{\alpha}$, we extracted the values of $H_{\mathrm{c2}}(0)$, $T_{\mathrm{c}}$, and $\alpha$ at pressures of 2.17~GPa, 1.35~GPa, and 0.68~GPa. The analysis reveals an increase in $H_{\mathrm{c2}}(0)$ and a decrease in $\alpha$ with increasing pressure, suggesting that 1T-MoTe$_2$ enters a strong-coupling superconducting regime under high pressure.
    
    \item The structural phase transition from T$\mathrm{_d}$ to 1T$^{\prime}$ phase was not detected through either the temperature or pressure dependence of the $^{125}$Te NMR line profiles. This indicates that the transferred hyperfine interactions are largely insensitive to changes in crystal symmetry and slight variations in lattice parameters.
    
    \item The temperature and pressure dependence of $1/T_1T$ is consistent with the prediction from the DFT calculations. The relative DOS, estimated from the Korringa relation, was found to increase with pressure up to approximately 0.7~GPa. This mirrors the pressure dependence of the $T_\mathrm c$, which can be explained by the conventional BCS theory. In contrast, the DOS gradually decreases with increasing pressure above 0.7~GPa, while $T_c$ continues to increase. This discrepancy suggests that additional contributions -- likely of magnetic origin -- play a significant role in the emergence of superconductivity under higher pressure.
    
    \item In the superconducting state, a multi-gap superconducting behavior was observed at 2.17~GPa and 3~K, where the system remains in the topologically trivial 1$T'$ phase. 
    
    \end{itemize}

The present study on 1T-MoTe$_2$ highlights that many open questions remain concerning the relationship between electronic topology and superconductivity. In particular, NMR data—especially the temperature dependence of $1/T_1T$ in the superconducting state—is highly anticipated to provide further insight. A detailed investigation of the pressure and temperature dependence of the superconducting order parameter is essential for exploring the potential realization of topological superconductivity. We believe that the current work offers valuable progress and sheds light on this promising direction.

\section*{Acknowledgment}
We acknowledge fruitful discussions with Mario Moda Piva, Deepa Kashinathan and Michael Nicklas. We would like to thank Vicky Hasse for support during the synthesis. Work at Brookhaven National Laboratory was supported by the U.S. Department of Energy, Office of Science, Office of Basic Energy Sciences under Contract No. DE-SC0012704 (materials synthesis). T. F. appreciate the financial support from JSPS KAKENHI Grants (No. 15K21732).

\section*{Present address}  
\begin{itemize}
\item[$\star$]Department of Physics, Indian Institute of Technology Palakkad, Kerala 678623, India
\item[$\dagger$]Center for Correlated Matter and School of Physics, Zhejiang University, Hangzhou 310058, China
\item[*]Shanghai Key Laboratory of Material Frontiers Research in Extreme Environments
(MFree), Shanghai Advanced Research in Physical Sciences (SHARPS), Shanghai 201203,
China

\end{itemize}

\bibliography{reference}

@article{xu2015,
  title={Discovery of a Weyl fermion semimetal and topological Fermi arcs},
  author={Xu, Su-Yang and Belopolski, Ilya and Alidoust, Nasser and Neupane, Madhab and Bian, Guang and Zhang, Chenglong and Sankar, Raman and Chang, Guoqing and Yuan, Zhujun and Lee, Chi-Cheng and others},
  journal={Science},
  volume={349},
  number={6248},
  pages={613--617},
  year={2015},
  publisher={American Association for the Advancement of Science}
}

@article{lv2015,
  title={Experimental discovery of Weyl semimetal TaAs},
  author={Lv, BQ and Weng, HM and Fu, BB and Wang, X Ps and Miao, Hu and Ma, Junzhang and Richard, P and Huang, XC and Zhao, LX and Chen, GF and others},
  journal={Physical Review X},
  volume={5},
  number={3},
  pages={031013},
  year={2015},
  publisher={APS}
}

@article{huang2015,
  title={Observation of the chiral-anomaly-induced negative magnetoresistance in 3D Weyl semimetal TaAs},
  author={Huang, Xiaochun and Zhao, Lingxiao and Long, Yujia and Wang, Peipei and Chen, Dong and Yang, Zhanhai and Liang, Hui and Xue, Mianqi and Weng, Hongming and Fang, Zhong and others},
  journal={Physical Review X},
  volume={5},
  number={3},
  pages={031023},
  year={2015},
  publisher={APS}
}

@article{zhang2016,
  title={Signatures of the Adler--Bell--Jackiw chiral anomaly in a Weyl fermion semimetal},
  author={Zhang, Cheng-Long and Xu, Su-Yang and Belopolski, Ilya and Yuan, Zhujun and Lin, Ziquan and Tong, Bingbing and Bian, Guang and Alidoust, Nasser and Lee, Chi-Cheng and Huang, Shin-Ming and others},
  journal={Nature communications},
  volume={7},
  number={1},
  pages={1--9},
  year={2016},
  publisher={Nature Publishing Group}
}

@article{sankar2017,
  title={Polymorphic layered MoTe2 from semiconductor, topological insulator, to Weyl semimetal},
  author={Sankar, Raman and Narsinga Rao, G and Muthuselvam, I Panneer and Butler, Christopher and Kumar, Nitesh and Senthil Murugan, G and Shekhar, Chandra and Chang, Tay-Rong and Wen, Cheng-Yen and Chen, Chun-Wei and others},
  journal={Chemistry of Materials},
  volume={29},
  number={2},
  pages={699--707},
  year={2017},
  publisher={ACS Publications}
}

@article{2016YQ,
  author = {Qi, Y. and Naumov, P. G. and Ali, M. N. and Rajamathi, C. R. and Schnelle, W. and Barkalov, O. and Hanfland, M. and Wu, S. C. and Shekhar, C. and Sun, Y. and S\"u{\ss}, V. and Schmidt, M. and Schwarz, U. and Pippel, E. and Werner, P. and Hillebrand, R. and F\"orster, T. and Kampert, E. and Parkin, S. and Cava, R. J. and Felser, C. and Yan, B. and Medvedev, S. A.},
  journal = {Nat. Commun.},
  volume = {7},
  pages = {11038},
  year = {2016}
}

@article{Hu2019a,
  author = {Hu, Y. J. and Chan, Y. T. and Lai, K. T. and Ho, K. O. and Guo, X. and Sun, H.-P. and Yip, K. Y. and Ng, D. H. L. and Lu, H.-Z. and Goh, S. K.},
  journal = {Phys. Rev. Mater.},
  volume = {3},
  pages = {034201},
  year = {2019}
}

@article{Takahashi2017,
  author = {Takahashi, H. and Akiba, T. and Imura, K. and Shiino, T. and Deguchi, K. and Sato, N. K. and Sakai, H. and Bahramy, M. S. and Ishiwata, S.},
  journal = {Phys. Rev. B},
  volume = {95},
  pages = {100501},
  year = {2017}
}

@article{Piva2023,
  author = {Piva, M. M. and Kutelak, L. O. and Borth, R. and Liu, Y. and Petrovic, C. and dos Reis, R. D. and Nicklas, M.},
  journal = {Phys. Rev. Mater.},
  volume = {7},
  pages = {L111801},
  year = {2023}
}

@article{Heikes2018,
  author = {Heikes, C. and Liu, I.-L. and Metz, T. and Eckberg, C. and Neves, P. and Wu, Y. and Hung, L. and Piccoli, P. and Cao, H. and Leao, J. and Paglione, J. and Yildirim, T. and Butch, N. P. and Ratcliff, W.},
  journal = {Phys. Rev. Mater.},
  volume = {2},
  pages = {074202},
  year = {2018}
}

@article{Guguchia2017,
  author = {Guguchia, Z. and von Rohr, F. and Shermadini, Z. and Lee, A. T. and Banerjee, S. and Wieteska, A. R. and Marianetti, C. A. and Frandsen, B. A. and Luetkens, H. and Gong, Z. and Cheung, S. C. and Baines, C. and Shengelaya, A. and Taniashvili, G. and Pasupathy, A. N. and Morenzoni, E. and Billinge, S. J. L. and Amato, A. and Cava, R. J. and Khasanov, R. and Uemura, Y. J.},
  journal = {Nat. Commun.},
  volume = {8},
  pages = {1082},
  year = {2017}
}

@article{hang2024,
  author = {Zhang, D. and Xu, Z. and Le, T. and Chen, C. and Ye, G. and Shi, F. and Luo, S. and Shi, Y. and Lu, X.},
  journal = {Phys. Rev. B.},
  volume = {109},
  pages = {144506},
  year = {2024}
}

@article{Koepernik1999,
  author = {Koepernik, K. and Eschrig, H.},
  journal = {Phys. Rev. B},
  volume = {59},
  pages = {1743},
  year = {1999}
}

@article{Tran2009,
  author = {Tran, F. and Blaha, P.},
  journal = {Phys. Rev. Lett.},
  volume = {102},
  pages = {226401},
  year = {2009}
}

@article{Perdew1992,
  author = {Perdew, J. P. and Wang, Y.},
  journal = {Phys. Rev. B},
  volume = {45},
  pages = {13244},
  year = {1992}
}

@article{Wang2016,
  author = {Wang, Z. and Gresch, D. and Soluyanov, A. A. and Xie, W. and Kushwaha, S. and Dai, X. and Troyer, M. and Cava, R. J. and Bernevig, B. A.},
  journal = {Phys. Rev. Lett.},
  volume = {117},
  pages = {056805},
  year = {2016}
}

@article{Yang2017,
  author = {Yang, J. and Colen, J. and Liu, J. and Nguyen, M. C. and Chern, G.-W. and Louca, D.},
  journal = {Sci. Adv.},
  volume = {3},
  pages = {eaao4949},
  year = {2017}
}

@article{Werthamer1966,
  author = {Werthamer, N. R. and Helfand, E. and Hohenberg, P. C.},
  journal = {Phys. Rev.},
  volume = {147},
  pages = {295},
  year = {1966}
}

@article{Muller2001,
  author = {M\"uller, K.-H. and Fuchs, G. and Handstein, A. and Nenkov, K. and Narozhnyi, V. N. and Eckert, D.},
  journal = {J. Alloys Compd.},
  volume = {322},
  pages = {L10},
  year = {2001}
}

@article{Narath1968,
  author = {Narath, H. and Weaver, H.},
  journal = {Phys. Rev.},
  volume = {175},
  pages = {373},
  year = {1968}
}

@article{Moriya1963,
  author = {Moriya, T.},
  journal = {J. Phys. Soc. Jpn.},
  volume = {18},
  pages = {516},
  year = {1963}
}

@book{Slichter1990,
  author = {Slichter, C. P.},
  publisher = {Springer},
  address = {New York},
  edition = {3rd},
  year = {1990}
}

@article{Korringa1950P,
  author = {Korringa, J.},
  journal = {Physica},
  volume = {16},
  pages = {601},
  year = {1950}
}

@article{Bardeen1957,
  author = {Bardeen, J. and Cooper, L. N. and Schrieffer, J. R.},
  journal = {Phys. Rev.},
  volume = {108},
  pages = {1175},
  year = {1957}
}

 \end{document}